\documentclass[12pt]{article}
\usepackage[dvipdfmx]{graphicx} 
\usepackage{hyperref}
\usepackage{amsmath, amssymb, bm, amsthm, mathrsfs, braket} 
\usepackage{color} 

\newcommand{\del}{\partial} 
\newcommand{\lap}{\triangle} 
\newcommand{\abs}[1]{\left|#1\right|} 
\renewcommand{\Lambda}{\varLambda} 

\newcommand{\calO}{\mathcal{O}}
\newcommand{\ii}{{\rm i}}

\newcommand{\dd}{{\rm d}}
\renewcommand{\ii}{{\rm i}}
\makeatletter

\@addtoreset{equation}{section}
\makeatother

\begin{document}

\begin{titlepage}

\setcounter{page}{0}

\renewcommand{\thefootnote}{\fnsymbol{footnote}}

\begin{flushright}
YITP-21-05, UT-Komaba/21-2
\end{flushright}

\vskip 1.35cm

\begin{center}
{\Large \bf 
A Note on Commutation Relation 
in Conformal Field Theory
}

\vskip 1.2cm 

{\normalsize
Lento Nagano${}^{a}$~\footnote{lento.nagano(at)gmail.com} and Seiji Terashima${}^{b}$~\footnote{terasima(at)yukawa.kyoto-u.ac.jp}
}

\vskip 0.8cm

${}^{a}${\it
Institute of Physics, University of Tokyo, Komaba,
\\
Meguro-ku, Tokyo 153-8902, Japan
}

${}^{b}${\it
Yukawa Institute for Theoretical Physics, Kyoto University, 
\\
Kyoto 606-8502, Japan
}

\end{center}

\vspace{12mm}

\centerline{{\bf Abstract}}

In this note, we explicitly
compute the vacuum expectation value of the commutator of scalar fields
in a $d$-dimensional conformal field theory on the cylinder.
We find from explicit calculations that we need smearing not only in space but also in time to have finite commutators except for those of free scalar operators. 
Thus
the equal time commutators of the scalar fields 
are not well-defined for a non-free conformal field theory, even if which is defined from the Lagrangian.
We also have the commutator for a conformal field theory on Minkowski space, instead of the cylinder, by taking the small distance limit.
For the conformal field theory on Minkowski space, the above statements are also applied.

\end{titlepage}
\newpage

\tableofcontents
\vskip 1.2cm 

\section{Introduction and summary}

The commutation relation or commutator of the quantum fields
is the fundamental objects in quantum field theory (QFT)
in the operator formalism.
Indeed, usually the QFT is given by the canonical commutator,
which is defined at a fixed  time slice, of 
fundamental fields and the Hamiltonian, which describes the time evolution, in the operator formalism.
Even for the theory without the canonical commutator,
the commutator is important.
For the two dimensional conformal field theory (CFT),
the Virasoro algebra and the current algebra, which 
are the equal time commutators, play the important roles.
These can be derived from the operator product expansion of fields
by the contour integrals using the infinitely many current conservation laws.

Higher dimensional ($d\geq3$) CFTs are also very important in theoretical physics, such as condensed matter physics and AdS/CFT correspondence~\cite{Maldacena}, and so on. 
Recently they have been significantly studied by the conformal bootstrap since a seminal work~\cite{Rattazzi:2008pe}.~\footnote{See~\cite{Ry, SD} for reviews on this topic.}
However, for the $d$-dimensional CFT (or general fields in 2d CFT),
the commutators have not been studied intensively partly because 
there are not infinitely many conserved current.~\footnote{
Constraints on commutators and their application were investigated by recent works such as~\cite{Hartman:2015lfa,Hartman:2016dxc,Hartman:2016lgu,Afkhami-Jeddi:2016ntf,Kulaxizi:2017ixa}.
The stress tensor commutators 
were studied in old works, {\it e.g.}~\cite{Schwinger:1962wd,Schwinger:1963zz,Deser:1967zzf}
and also in recent works~\cite{Huang:2019fog,Huang:2020ycs}.}
On the other hand, the commutator of fields in the cylinder $\mathbb{R} \times \mathbb{S}^{d-1}$ 
can be derived 
from the operator product expansion (OPE) for the higher dimensional CFT recently~\cite{Terashima:2019wed}. 
In particular, the vacuum expectation value (VEV) of 
the commutator is determined by the most singular part of the OPE,
which is essentially the two point function.

In this note, we explicitly
compute the VEV of the commutator of (primary) scalar fields
in $d$-dimensional CFT on the cylinder for $d \geq 2$.
The commutators are expressed by an infinite summation.
We observe a difference between the commutators for a free CFT and a non-free CFT as follows.
\begin{itemize}
\item
For the commutators of free primary scalar fields, 
we only need to smear operators in space and don't need to do so in time to have finite values. So the equal time commutators of operators smeared in space are well-defined.
\item
For the commutators of primary scalar fields in a non-free CFT,
we need to smear operators not only in space but also in time to have finite values. So the equal time commutators of smeared operators are ill-defined.\footnote{
The two-point function is well-defined and finite
except on the light cone where it diverges.
What we will discuss is 
the property of this divergence for a non-trivial CFT, by which we can
see that the equal time commutators are ill-defined for it.
}
\end{itemize}
The latter fact is related to the fact that the weight of the 
K\"{a}ll\'{e}n-Lehmann-like representation for 
the CFT is not normalizable.
We also have the commutator for the CFT on Minkowski space, instead of the cylinder, by taking the small distance limit.
Besides we can explicitly perform a summation when $\Delta=d/2$.
We hope our results will be useful for future studies of 
the CFT, in particular for the AdS/CFT correspondence in  
the operator formalism~\cite{Terashima:2019wed,Terashima:2017gmc, Terashima:2020uqu}.

We will also argue that the quantum field theory with a non-trivial UV fixed point has divergent equal time commutators.
On the other hand, an asymptotic free quantum field theory has
a well-defined equal time commutators as described.

This paper is organized as follows. In Section~\ref{sec:ope} we review the OPE
in general CFTs and explain how to compute the commutator from the OPE.
In section~\ref{sec:comm}, we compute the VEV of the commutators of the scalar fields in CFTs.
First we evaluate them for a free CFT in Section~\ref{sec:free-cft}, and then discuss them for a non-free CFT in Section~\ref{sec:gen-cft}.

\section{OPE}\label{sec:ope}
In this section we review the OPE of primary scalar fields.
Let us consider a scalar primary operator $\mathcal{O}$ with a conformal dimension $\Delta$.
In general, the OPE between two operators in Euclidean flat space is given by 
\begin{equation}\label{eq:general-ope}
	\mathcal{O}(x_{1})\mathcal{O}(x_{2})
	=
	\sum_{\mathcal{O}_{p}: \text{primary}}
		C_{\mathcal{O}\mathcal{O}\mathcal{O}_{p}}
		f_{\mu_{1}\dots\mu_{{l}_{p}}}(x_{12},\del_{2})
		\mathcal{O}_{p}^{\mu_{1}\dots\mu_{{l}_{p}}} (x_2) \,,
\end{equation}
where $x_{12}=|x_1-x_2|$ and 
$f_{\mu_{1}\dots\mu_{{l}_{p}}}(x_{12},\del_{2})$ is a function which can be determined only by the representation theory of the conformal symmetry.
The most singular term in \eqref{eq:general-ope} is the contribution from identity operator~\footnote{We mean "taking most singular part" by $\sim$.},
\begin{equation}\label{eq:general-ope2}
	\mathcal{O}(x_{1})\mathcal{O}(x_{2})
	\sim
	x_{12}^{-2\Delta} \, ,
\end{equation}
where we normalized $\mathcal{O}(x_{1})$ usually.
We focus on a contribution from this term.
Note that if we consider OPE in a two point function, only an identity term contribute,
\begin{equation}\label{eq:general-ope3}
	\Braket{\mathcal{O}(x_{1})\mathcal{O}(x_{2})}
	=
	x_{12}^{-2\Delta} \,,
\end{equation}
since the one-point functions of any operators except for an identity vanish on a conformaly flat manifold.

\subsection{OPE in cylinder coordinates}

First we parametrize a position in Euclidean flat space $\mathbb{R}^{d}$ by $x^{\mu}=r \, e^{\mu}(\Omega)$, 
where $r=\abs{x}$ and $e^{\mu}(\Omega)$ is a unit vector, i.e. $e^{\mu}(\Omega) e_{\mu}(\Omega)=1$. We parametrize unit vector $e^{\mu}$ by angular variables $\Omega$.
Then we move to the cylinder coordinates via 
\begin{align}
r=e^{\tau}.
\end{align}
Operators which live in the cylinder coordinates are denoted by $\mathcal{O}^{\text{cyl}}$ and they are related to the corresponding operators in flat space by
\begin{equation}
	\mathcal{O}^{\text{cyl}}(\tau,\Omega)
	=
	r^{\Delta}\mathcal{O}(r,\Omega) \,.
\end{equation}
In the cylinder coordinates, the OPE can be written as
\begin{equation}
\mathcal{O}^{\text{cyl}}(\tau_{1},\Omega_{1})
\mathcal{O}^{\text{cyl}}(\tau_{2},\Omega_{2})
\sim
r_{1}^{\Delta}r_{2}^{\Delta}x_{12}^{-2\Delta} \,.
\end{equation}
We suppress a superscript ``cyl'' below.
We can expand $x_{12}^{-2\Delta}$ as follows~\cite{Terashima:2019wed}.
\begin{align}
x_{12}^{-2\Delta}
&=
\frac{1}{r_{>}^{2\Delta}}
\sum_{s=0}^{\infty}
\left(
\frac{r_{<}}{r_{>}}
\right)^{s}
\sum_{n=0}^{[s/2]}
\left(d^{\Delta}\right)_{s}^{s-2n}
\sum_{m}
Y_{s-2n,m}(\Omega_{1})
Y_{s-2n,m}(\Omega_{2})
\\
&=
\frac{1}{r_{>}^{2\Delta}}
\sum_{n=0}^{\infty}
\sum_{l=0}^{\infty}
\left(
\frac{r_{<}}{r_{>}}
\right)^{2n+l}
\left(d^{\Delta}\right)_{2n+l}^{l}
\sum_{m}
Y_{l,m}(\Omega_{1})
Y_{l,m}(\Omega_{2}) ,
\label{ex1}
\end{align}
where $r_{>}$ and $r_{<}$ are the larger and smaller ones of $|x_1|$ and $|x_2|$, respectively and
\begin{equation}
\left(d^{\Delta}\right)_{2n+l}^{l}
=
\frac{2\pi^{d/2} \,
\Gamma(\Delta+n+l)\Gamma(\Delta+1-d/2+n)}{\Gamma(\Delta)\Gamma(\Delta+1-d/2)\Gamma(n+1)\Gamma(n+l+d/2)} .
\end{equation}
For the normalization of the spherical harmonics, see the appendix~\ref{app:spherical}.
Thus, 
we have
\begin{align}
&\mathcal{O}(\tau_{1},\Omega_{1})
\mathcal{O}(\tau_{2},\Omega_{2})
\notag
\\
&\sim
\sum_{n,l=0}^{\infty}
e^{-(\Delta+2n+l) |\tau_{12}| }\left(d^{\Delta}\right)_{2n+l}^{l}
\sum_{m}
Y_{l,m}(\Omega_{1})
Y_{l,m}(\Omega_{2})
\\
&=
\sum_{n,l=0}^{\infty}
e^{-(\Delta+2n+l) |\tau_{12}| }\left(d^{\Delta}\right)_{2n+l}^{l} \tilde{C}_{l}(\Omega_{12}),
\end{align}
where $\tau_{12}:=\tau_{1}-\tau_{2}$, $\Omega_{12}:=e^{\mu}(\Omega_1) e_{\mu}(\Omega_2)=\cos \theta_{12}$, where
$\theta_{12}$ is the angle between the two points in $\mathbb{S}^{d-1}$,
and
\begin{equation}
\tilde{C}_{l}(\Omega_{12}):=\sum_{m}
Y_{l,m}(\Omega_{1})
Y_{l,m}(\Omega_{2})
=
\frac{d+2 l-2  }{d-2} 
C^{d/2-1}_{l}(\Omega_{12}),
\end{equation}
where $C^{\alpha}_{l} (x)$ is the Gegenbauer polynomial~\cite{Avery}.~\footnote{
For $d=2$, $\sum_{m}
Y_{l,m}(\Omega_{1}^{\mu})
Y_{l,m}(\Omega_{2}^{\mu})
=\frac{2\pi^{d/2}}{\Gamma(d/2)} (2 \cos (l \theta_{12})- \delta_{l, 0})$.
}
For the normalization of the Gegenbauer polynomials, see Appendix~\ref{sec:gegenbauer}.

We have considered only the most singular part of the OPE, however,
other parts also can be expanded in the same way.
Below, we continue dealing only with the most singular part.
For other parts, we can also compute the commutator formally~\cite{Terashima:2019wed}
although we need the explicit OPE data to give an explicit result.

\section{Commutation relation}
\label{sec:comm}

\subsection{Ordering of the operators}

In the previous section, 
the OPE (\ref{eq:general-ope})  is regarded as the expansion 
in the correlation function, where the fields, which are regarded as the path-integral variables, can commute each other
and the ordering is not relevant.
Below, 
we will consider the CFT in the operator formalism
and regard the fields in the 
OPE (\ref{eq:general-ope}) as operators acting on the Hilbert space.
More precisely, we will regard an operator $\mathcal{O}_{E}(\tau,\Omega)$
which is the field in the Euclidean cylinder as the operator acting on the Hilbert space
of  the CFT${}_d$ on the sphere~$\mathbb{S}^{d-1}$.~\footnote{Below we denote the operator whose argument is an Euclidean time by $\mathcal{O}_{E}$.}
Then, 
the operator ordering corresponding to the correlation function with the condition~$\tau_{1}>\cdots>\tau_{n}$~\footnote{
For a quantum mechanics (with a finite degrees of freedom),
we do not need to expand like (\ref{ex1}) and we can consider
any ordering of the operators. 
In quantum field theories, the operators which is not ordered as (\ref{ord1}) 
have a diverging expectation values.
However, the local field should be smeared for a finite expectation value 
and then the smeared operators which is not ordered as (\ref{ord1}),
where smearing region is small compared with the time distances
will have finite values.
In this sense, the other orderings are possible. 
}
is as follows,
\begin{equation}
\mathcal{O}_{E}(\tau_{1},\Omega_{1})
\mathcal{O}_{E}(\tau_{2},\Omega_{2})\cdots
\mathcal{O}_{E}(\tau_{n},\Omega_{n}),
\quad (\tau_{1}>\cdots>\tau_{n}),
\label{ord1}
\end{equation}
where $\mathcal{O}_{E}(\tau,\Omega) =e^{ \tau H} \mathcal{O}_{E} (0,\Omega) e^{ -\tau H}$
and $H$ is the Hamiltonian (which is the dilatation operator).
For this ordering, we can apply the OPE of the two operators 
using the expansion (\ref{ex1}).

We can define the the product of the operators for a complex $\tau$ by 
the analytic continuation by the expansion (\ref{ex1}) for
\begin{equation}
\mathcal{O}_{E}(\tau_{1},\Omega_{1})\mathcal{O}_{E}(\tau_{2},\Omega_{2})\cdots\mathcal{O}_{E}(\tau_{n},\Omega_{n}),
\quad ({\rm Re} (\tau_{1}) >\cdots>{\rm Re} (\tau_{n}) ),
\label{ord2}
\end{equation}

When we want to move to a CFT in Lorentzian signature, 
we evolve a Lorenzian time with the ordering given by the small Euclidean time $\epsilon_{i}$ fixed, and then take $\epsilon_{i}\to0$ limit:
\begin{align}
&\mathcal{O}_{L}(t_{1},\Omega_{1})\mathcal{O}_{L}(t_{2},\Omega_{2})\cdots\mathcal{O}_{L}(t_{n},\Omega_{n})
\notag\\
&:=
\lim_{\epsilon_i \rightarrow 0}
\mathcal{O}_{E}(\tau_{1}=\epsilon_{1}+\ii t_{1},\Omega_{1})\cdots\mathcal{O}_{E}(\tau_{n}=\epsilon_{n}+\ii t_{n},\Omega_{n})
\quad (\epsilon_{1}>\cdots>\epsilon_{n}) ,
\end{align}
where we denote operators whose argument is a Lorentzian time as $\mathcal{O}_{L}(t,\Omega)$.~\footnote{
In operator formalism, the Hilbert space and operators acting on it  
are same for Lorentzian and Euclidean signature.
``Lorentzian'' means that the operator is evolved by $e^{\ii H t}$, instead of 
$e^{H \tau }$ for Euclidean case.
}

Note that the local operators has a diverging expectation value, thus
we need to consider the smearing (or distributions) of them.



\subsection{Free CFT}
\label{sec:free-cft}

First, we consider the free massless scalar 
which means 
$\Delta=\frac{d}{2}-1$, as a simplest example.
For this case, we have
\begin{align}
\left(d^{\text{free}}\right)_{2n+l}^{l}&:=\left(d^{\Delta=d/2-1}\right)_{2n+l}^{l}\\
&=
\frac{2\pi^{d/2}}{\Gamma(d/2)} \frac{\Gamma(d/2)\Gamma(d/2-1+n+l)\Gamma(n)}{\Gamma(d/2-1)\Gamma(0)\Gamma(n+1)\Gamma(n+l+d/2)} 
\\
&=
\delta_{n,0}\frac{d/2-1}{(l+d/2-1)}  \frac{2\pi^{d/2}}{\Gamma(d/2)} 
\\
&=
\delta_{n,0}\frac{\Delta}{(\Delta+l)}\frac{2\pi^{d/2}}{\Gamma(d/2)} 
.
\end{align}
With this, we can write down the (singular part of) OPE as
\begin{align}
\mathcal{O}_{E} (\tau_{1},\Omega_{1})
\mathcal{O}_{E} (\tau_{2},\Omega_{2})
\sim
\frac{2\pi^{d/2}\Delta}{\Gamma(d/2)}
\sum_{l=0}^{\infty}
\frac{e^{-(\Delta+l)\tau_{12}}}{(\Delta+l)}\tilde{C}_{l}(\Omega_{12}) ,
\end{align}
where we assume  ${\rm Re} (\tau_{1}-\tau_{2})>0$. 
Using this, the VEV of the commutation relation of the two local operators
is computed as 
$\bra{0} [\mathcal{O}_L(t_{1},\Omega_{1}),  \mathcal{O}_L (t_{2},\Omega_{2})] \ket{0}
=\lim_{\epsilon \rightarrow 0} A_{\epsilon}(t_{12},\Omega_{12}) $
where
\begin{align}
A_{\epsilon}(t_{12},\Omega_{12})
&:=\bra{0} 
\mathcal{O}_{E}(\tau_{1}=\epsilon+\ii t_{1},\Omega_{1})
\mathcal{O}_{E}(\tau_{2}=\ii t_{2},\Omega_{2})
\notag
\\
&\qquad
-
\mathcal{O}_{E}(\tau_{2}=\ii t_{2},\Omega_{2})
\mathcal{O}_{E}(\tau_{1}=-\epsilon+\ii t_{1},\Omega_{1})\ket{0}
\\
&=\bra{0} 
\mathcal{O}_{L}(t_{1}-\ii \epsilon,\Omega_{1})
\mathcal{O}_{L}(t_{2},\Omega_{2})
\notag
\\
&\qquad
-
\mathcal{O}_{L}(t_{2},\Omega_{2})
\mathcal{O}_{L}(t_{1}+\ii \epsilon,\Omega_{1})\ket{0}
\\
&=
-2 \ii \frac{2\pi^{d/2}}{\Gamma(d/2)}  \sum_{l=0}^{\infty} \Delta
\frac{e^{-(\Delta+l)\epsilon}\sin\left((\Delta+l)t_{12}\right)}{(\Delta+l)} \tilde{C}_{l}(\Omega_{12}) ,
\label{eq:commutation-oo-free}
\end{align}
and we assume  $\epsilon>0$ and $t_{12}=t_{1}-t_{2}$.
Note that for the free theory, we know the commutator contains only the identity operator.
This means $ [\mathcal{O}_L(t_{1},\Omega_{1}),  \mathcal{O}_L (t_{2},\Omega_{2})] =
\lim_{\epsilon \rightarrow 0} A_{\epsilon}(t_{12},\Omega_{12}) $, where the identity operator is not explicitly written.

Precisely speaking, we need to take the limit after
the smearing of the local operators. Here we just need a space smearing, not a spacetime smearing.

We can easily check that this formally satisfies the equations of motion of the free field (which has the conformal mass term $\Delta^2$ on $\mathbb{S}^{d-1}$) as
\begin{equation}
\left( \frac{\partial^2}{\partial t_{1}^2}-  \lap_{\mathbb{S}^{d-1}(\Omega_1)}+\Delta^2 \right) [\mathcal{O}_{L}(t_{1},\Omega_{1}),  \mathcal{O}_{L}(t_{2},\Omega_{2})]=0,
\end{equation}
where we assume $t_1 \neq t_2$ or $\Omega_1 \neq \Omega_2$.
Here,  $\lap_{\mathbb{S}^{d-1}(\Omega_1)}$ is the Laplacian acting on $\Omega_1$
and satisfies
\begin{align}
 \lap_{\mathbb{S}^{d-1}} Y_{l,m}(\Omega)
=-l(l+d-2) Y_{l,m}(\Omega).
\end{align}

\subsubsection{Equal time commutator}

The equal time commutation relation can be easily computed as
\begin{equation}
[\mathcal{O}_{L}(t_{1},\Omega_{1}),  \mathcal{O}_{L}(t_{1},\Omega_{2})]=0
\end{equation}
because of the symmetry. 
Instead if we consider $[\dot{\calO},\calO]$, then we find 
\begin{align}
&
\bra{0}
\left[\frac{d}{dt_{1}} \mathcal{O}_{L}(t_{1},\Omega_{1}),  \mathcal{O}_{L}(t_{1},\Omega_{2})\right] \ket{0}
\notag
\\
&= 
\left.\frac{d}{dt_{1}}A_{\epsilon}(t_{12},\Omega_{12})\right|_{t_{12}=0}
\\
&=
 2 \ii \frac{2\pi^{d/2}}{\Gamma(d/2)}  \Delta
\sum_{l=0}^{\infty} e^{-(\Delta+l)\epsilon}
\sum_{m}
Y_{l,m}(\Omega_{1})
Y_{l,m}(\Omega_{2})
\\
& \rightarrow
2 \ii \frac{2\pi^{d/2}}{\Gamma(d/2)}  \Delta \,
\delta(\Omega_{1}-\Omega_{2}) 
\quad(\epsilon\to0),
\label{eq:etc-free}
\end{align}
where the expression in the final line is interpreted as a distribution.

\subsubsection{Flat space limit}

Finally, we will consider the commutator of very close two operators, {\it i.e.} the commutator with
$|t_{12}| \ll 1$ and $|\Omega_{12} -1| \ll 1$.
In this limit, we expect that the commutator becomes the one for 
the theory on Minkowski space which is given by the invariant Delta function.~\footnote{In this paper we call this limit as a flat space limit, but this is different from the usual flat space limit as in~\cite{Penedones:2010ue,Okuda:2010ym}.
}
For this, we will use the formula of the Gegenbaur polynomial as the Jacobi polynomial,
\begin{align}
C_n^\alpha(x)= \frac{\Gamma(2 \alpha+n)}{\Gamma(2 \alpha)} 
\frac{\Gamma(\alpha+1/2)}{\Gamma(\alpha+1/2+n)} 
P_n^{\alpha-1/2, \alpha-1/2} (x),
\end{align}
and the asymptotics of the Jacobi polynomials. Near the point $x=1$,
we have~\cite{asymptotic2}
\begin{align}
\lim_{n \rightarrow \infty} n^{-\alpha} P_n^{\alpha, \beta} (\cos (z/n)) = (z/2)^{-\alpha} J_\alpha (z),
\end{align}
where $J_\alpha(z)$ is the Bessel function of the first kind.
Using this, we can see that the commutator (\ref{eq:commutation-oo-free}) in the limit 
becomes 
\begin{align}
\lim_{\epsilon\to0}
A_{\epsilon}(t_{12},\Omega_{12})
& =
-2 \ii \frac{2\pi^{d/2}}{\Gamma(d/2)}  \sum_{l=0}^{\infty} \Delta
\frac{\sin\left((\Delta+l)t_{12}\right)}{(\Delta+l)} \tilde{C}_{l}(\Omega_{12})  \\
& = -2 \ii 
\frac{2\pi^{d/2}}{\Gamma(d/2)} 
\Delta \int_0^\infty d k \frac{1}{k}
\sin\left( k \, t \right)
\tilde{C}_{k/\varepsilon}(\cos(\varepsilon r)) +\cdots \\
&=  -\ii 
\frac{2\pi^{d/2}}{\Gamma(d/2)} 
\int_0^\infty d k 
\frac{\Gamma(d-1+k/\varepsilon)}{\Gamma(d-2)} 
\frac{\Gamma(d/2-1/2)}{\Gamma(d/2-1/2+k/\varepsilon)} 
\notag\\&\qquad\qquad\qquad\qquad\times
\frac{1}{k}
\sin\left( k \, t \right)
P^{\frac{d-3}{2},\frac{d-3}{2} }_{k/\varepsilon}(\cos(\varepsilon r)) +\cdots \\
& =  -\ii 
\frac{2\pi^{d/2}}{\Gamma(d/2)} 
\frac{\Gamma((d-1)/2)}{\Gamma(d-2)} 
\int_0^\infty d k \frac{1}{k}
\left( \frac{k}{\varepsilon} \right)^{d-2}
\left( \frac{ k r}{2} \right)^{-(d-3)/2}
\notag\\&\qquad\qquad\qquad\qquad\times
\sin\left( k \, t \right)
J_{(d-3)/2} (k r) +\cdots,
\label{eq:flat-space-limit-free}
\end{align}
where the large $l$ contributions are dominant, then
we replace $l=k/\varepsilon$, $t_{12}= \varepsilon t$ and $\Omega_{12}=\cos (\theta_{12})=\cos (\varepsilon r)$ and $\cdots$ means the terms suppressed in the small $\varepsilon$ limit.

On the other hand, the commutator of the free scalar theory with a mass $\mu$ on $d$-dimensional Minkowski space
is given by 
\begin{align}
[\phi(x),\phi(0)] = \frac{1}{(2 \pi)^{d-1}} \int \prod_{i=1}^{d-1} \dd k_i (2 \omega(k))^{-1} \,
( e^{-\ii \omega(k) t+\ii k_i x^i} - e^{\ii\omega(k) t-\ii k_i x^i})
\label{eq:flat-space-commutator-free}
\end{align}
where $\omega(k)=\sqrt{k_i k^i+\mu^2}$ and we took the standard normalization.~\footnote{Here we introduced the mass of the scalar field for later convenience. We will finally take $\mu=0$.}
We can integrate out the angular directions of $k_i$ with the 
the following formula for the expansion of the plain wave in $\mathbb{R}^{d-1}$
by the spherical harmonics~\cite{Avery}:
\begin{align}
e^{\ii k_{i} x^{i} }
&= 
\frac{2\pi^{(d-1)/2}}{\Gamma((d-1)/2)} 
 (d-3)!! \sum_{l=0}^\infty \ii ^l \, j_l^{d-1} (kr) \sum_m Y_{l,m} (\Omega'_k) Y_{l,m}(\Omega'), \nonumber \\
&=4\pi^{(d-1)/2}  \sum_{l=0}^\infty \ii ^l \, 
\frac{ 2^{(d-5)/2} J_{(d-3)/2+l} (k r)  }{ (kr)^{(d-3)/2}}
\sum_m Y_{l,m} (\Omega'_k) Y_{l,m}(\Omega'),
\end{align}
where $r=\sqrt{x^{i} x_{i}}$, $k=\sqrt{k^{i} k_{i}}$,
$\Omega'$ and $\Omega'_k$ are the angular variables for $x^{i}$ and $k_{i}$,
respectively and $j_l^d(z)$ is 
the hyper spherical Bessel function which is written as
\begin{align}
j_l^d (z) & := 
\frac{\Gamma(d/2-1) 2^{d/2-2} J_{d/2-1+l} (z)  }{(d-4)!! z^{d/2-1}}.
\end{align}
After the integration over the angular directions $\Omega'_k$ in 
the momentum space $\mathbb{R}^{d-1}$, the constant mode of the spherical harmonics remains
and we find that 
\begin{align}
[\phi(x),\phi(0)] = 
 -\ii 2\pi^{(d-1)/2} 
\int_0^\infty d k \frac{1}{k}
\left( k \right)^{d-2}
\left( \frac{ k r}{2} \right)^{-(d-3)/2}
\sin\left( k \, t \right)
J_{(d-3)/2} (k r),
\end{align}
where we took the massless case $\mu=0$.
This coincides with the commutator in the flat space limit~
(\ref{eq:flat-space-commutator-free}) up to a numerical factor and 
$(\varepsilon)^{d-2}$ which is the scaling factor for the two free scalar fields.
Note that the normalizations of ${\cal{O}}(t,\Omega)$ and $\phi(x)$
are different.
Thus, we confirmed that the commutator reproduces
the usual commutator of the free theory on Minkowski space in the small distance limit.

\subsection{General CFT}
\label{sec:gen-cft}
In this section, we will consider the general scalar primary field 
whose dimension is above the unitarity bound, {\it i.e.} $\Delta> d/2-1$.
The OPE is given by 
\begin{align}
&\mathcal{O}_{E}(\tau_{1},\Omega_{1}^{\mu})
\mathcal{O}_{E}(\tau_{2},\Omega_{2}^{\mu})
\notag
\\
&\sim
\sum_{n,l=0}^{\infty}
e^{-(\Delta+2n+l)\tau_{12}}\left(d^{\Delta}\right)_{2n+l}^{l}
\tilde{C}_{l}(\Omega_{12}),
\end{align}
for $\text{Re}(\tau_{1})>\text{Re}(\tau_{2})$.
Thus, the VEV of 
the commutation relation of the two local operators
is computed as 
$\bra{0} [\mathcal{O}_{L}(t_{1},\Omega_{1}),  \mathcal{O}_{L}(t_{2},\Omega_{2})] \ket{0}
=\lim_{\epsilon \rightarrow 0} A_{\epsilon}(t_{12},\Omega_{12}) $
where 
\begin{align}
A_{\epsilon}(t_{12},\Omega_{12})
&:=
\bra{0} (
\mathcal{O}_{E}(\tau_{1}=\epsilon+\ii t_{1},\Omega_{1})
\mathcal{O}_{E}(\tau_{2}=\ii t_{2},\Omega_{2})
\notag
\\
&\qquad
-
\mathcal{O}_{E}(\tau_{2}=\ii t_{2},\Omega_{2})
\mathcal{O}_{E}(\tau_{1}=-\epsilon+\ii t_{1},\Omega_{1}) ) \ket{0}
\\
&=
\sum_{n,l=0}^{\infty}
e^{-(\Delta+2n+l)\epsilon}(-2\ii )\sin\left((\Delta+2n+l)t_{12}\right)
\notag
\\
&\qquad\times
\left(d^{\Delta}\right)_{2n+l}^{l}
\tilde{C}_{l}(\Omega_{12}) ,
\end{align}

Note that because of the factor $e^{-(\Delta+2n+l)\epsilon}$,
the summations over $n,l$ will converge if $\epsilon >0$.~\footnote{The summation over $n,l$ can be written by the summation over $\omega=2n+l$ and $l$ then the summation over $l$ is truncated since $\left(d^{\Delta}\right)_{\omega}$ vanish unless $l\leq \omega$.
Besides, 
$\tilde{C}_{l}(\Omega_{12})$ and $\left(d^{\Delta}\right)_{\omega}^{l}$ are power functions of $\omega,l$ in the $\omega,l \rightarrow \infty$ limit, thus this summation converges when $\epsilon>0$.
}
In order to take $\epsilon \rightarrow 0$ limit, we need to smear the local operators first in general.
We will explain this issue in Section~\ref{sec:gaussian} .
\subsubsection{Equal time commutator}
\label{sec:etc-general}

Let us consider the equal time commutator.
Usually, the equal time commutator is defined on a time slice, thus it should be
the distribution, like the delta-function, in space, not in spacetime.
Indeed, in a free CFT the equal time commutator of the free fields 
can be written by the delta-function with respect to~$\Omega$ as~\eqref{eq:etc-free}.
However, we will see that the equal time commutator in a general CFT except for a free one is not
defined even after integrating over space.~\footnote{This fact is related to another fact that 
the expectation value of the energy of the local state smeared over the space
is divergent for the operator of CFT except the free fields and homomorphic field in 2d case~\cite{Terashima:2020uqu}.  
}
This can be easily seen by considering 
the commutator of the~$l=0$ mode,~$\mathcal{O}_0(t) := \int d \Omega \, \mathcal{O}_L (t,\Omega)$,
which is maximally smeared over space.
The equal time commutator of this operator and its derivative are given by
\begin{equation}
\bra{0} [\partial_t \mathcal{O}_0(t), \mathcal{O}_0(t) ] \ket{0}=
-2 \ii \sum_{n=0}^\infty (\Delta+2n) \left(d^{\Delta}\right)_{2n}^{0}
\sim 
\sum_{n=0}^\infty n^{2 \Delta-d+1},
\label{eq:divergence}
\end{equation}
which diverges for $\Delta > d/2-1$ which is satisfied for the unitary CFT except for a free one.~\footnote{When $\Delta = d/2-1$ the above equation does not hold since only the $n=0$ term can contribute to the summation, so the equal time commutator does not diverge in this case. This is consistent with the results in Section~\ref{sec:free-cft}.}
As we will see below, if we smear an operator over spacetime instead of smearing over space we have a finite result.~\footnote{Note that this is consistent with a fact that in the axiomatic quantum field theory only correlators smeared over spacetime are considered.}

Remind that local operators smeared over the space in a certain time slice 
should have a finite commutator if the equal time commutators of the original operators in this time slice are well-defined.
Thus, for a CFT except the free CFT,~\footnote{
The energy momentum tensor and the homomorphic currents of $2$d CFT also
has the finite equal time commutators because the energy of a state is proportional to
the absolute value of the (angular) momentum and then there 
are no summation over $n$.
} 
the equal time commutators can not be defined as clearly seen by this divergence
of the one for the maximally smeared local operators.
This might be surprising because there are non-trivial CFTs which will have the Lagrangians and can be
defined by the canonical commutation relations with Hamiltonians, 
for example, the $4$d $\mathcal{N}=4$ supersymmetric gauge theory.
However, the gauge invariant operators are composite operators which need 
the renormalization. Then, such operators will not have a finite equal time commutator.

Note that this divergence will due to the high energy behavior of the theory.
Thus, the quantum field theory with a non-trivial UV fixed point, {\it i.e.} the one defined by the 
renormalization flow from the fixed point, will has divergent equal time commutators 
of the distributions for the local fields.
On the other hand, an asymptotic free quantum field theory will have
a well-defined equal time commutators as described in the usual text book.
We also expect that above properties do not depend on a choice of the time slice
because the local properties do not depend on the choice.

\subsubsection{Gaussian smeared local operators}
\label{sec:gaussian}

To take $\epsilon \rightarrow 0$ limit and 
obtain the commutator in Lorentzian spacetime,
we need to smear the local operators over spacetime instead of smearing over space.
Here, as an example of an explicit smearing, we introduce the Gaussian smeared (over time direction) local operators as ~\footnote{
The small Euclidian time $\epsilon$ caused the smearing of the local operator, however,
it also specifies the ordering of the operators.
To compute the commutator, we need another smearing. 
}
\begin{equation}
\mathcal{O}(t,\Omega)_{\delta} 
\equiv \frac{1}{\sqrt{2 \pi} \delta} \int^{\infty}_{-\infty}  d \alpha e^{-\frac{\alpha^2}{2 \delta^2} }  
\mathcal{O}_L (t+\alpha,\Omega).
\end{equation}
The commutator of the Gaussian smeared local operators is given by
\begin{align}
& \bra{0} [\mathcal{O}(t_1,\Omega_1)_{\delta}  ,\mathcal{O}(t_2,\Omega_2)_{\delta}  \ket{0} \\
= &
-\frac{2 \ii}{2 \pi \delta^2}  \sum_{n,l} 
\int^{\infty}_{-\infty}  d \alpha_1 d \alpha_2 e^{-\frac{(\alpha_1)^2+ (\alpha_2)^2}{2 \delta^2} }  
\notag\\&\quad\times
\sin( (\Delta+2n+l) t_{12}+\alpha_1-\alpha_2)
\left(d^{\Delta}\right)_{2n+l}^{l} \tilde{C}_{l}(\Omega_{12}) \\
= & -2 \ii \sum_{n,l} e^{- \delta^2 (\Delta+2 n+l)^2}  \sin( (\Delta+2n+l) t_{12})
\left(d^{\Delta}\right)_{2n+l}^{l} \tilde{C}_{l}(\Omega_{12}) ,
\end{align}
where we took the $\epsilon \rightarrow 0$ limit.
Then
the summations over $n,l$ converge even after taking $\epsilon\to0$
because of the Gasussian factor $e^{- \delta^2 (\Delta+2 n+l)^2}$.

Note that for $t_{12}=0$ the commutator is zero because of the 
symmetry~(and locality).

\subsubsection{Flat space limit}

If we only consider a small region in the spacetime (=the cylinder) ,
the theory is expected to become the CFT on Minkowski space.
We will see this for the commutators below.
Let us consider the following limit where $ \varepsilon \rightarrow 0$ and $t,r$ are fixed finite:
\begin{align}
t_{12}= \varepsilon t, \,\,\,\, \Omega_{12}(=\cos \theta_{12})= 1-(\varepsilon r)^2/2.
\end{align}
By defining
\begin{equation}
\left(d^{\Delta}\right)_{\infty}
:=
\frac{2\pi^{d/2}}{\Gamma(d/2)}
\frac{\Gamma(d/2)}{\Gamma(\Delta)\Gamma(\Delta+1-d/2)} ,
\end{equation}
the commutator in this limit becomes~\footnote{For more details on the approximation here, see Appendix~\ref{sec:approx}.}
\begin{align}
A_{\epsilon}(t_{12},\Omega_{12})
& =
\sum_{n,l=0}^{\infty}
e^{-(\Delta+2n+l)\epsilon}(-2\ii)\sin\left((\Delta+2n+l)t_{12}\right) \left(d^{\Delta}\right)_{2n+l}^{l} 
\tilde{C}_{l}(\Omega_{12}) \\
& = \left(d^{\Delta}\right)_{\infty}
\sum_{n,l=0}^{\infty}
e^{-(2n+l) } (-2\ii)
\sin\left((2n+l)t_{12}\right) 
\notag\\&\qquad\qquad\qquad\qquad\times
(n(n+l))^{\Delta-d/2}
\tilde{C}_{l}(\Omega_{12})
+\cdots
 \\
& = -\left(d^{\Delta}\right)_{\infty} \frac{\ii}{2 \varepsilon^2} \int_0^\infty d \mu \int_0^\infty d k
e^{-\sqrt{\mu^2+k^2} \frac{\epsilon}{\varepsilon} }
\sin\left( \sqrt{\mu^2+k^2} \, t \right)
\notag\\&\qquad\qquad\qquad\qquad\times
\left( \frac{\mu^2}{4 \varepsilon^2} \right)^{\Delta-d/2}
\tilde{C}_{k/\varepsilon}(\cos(\varepsilon r))
+\cdots
,
\end{align}
where contributions from $n,l \gg 1$ are dominant in the summations over $n,l$ and we defined 
\begin{align}
k=\varepsilon l, \,\,\, \mu^2 =\varepsilon^2 ((2n+l)^2-l^2),
\end{align}
which are considered as continuous variables.
Thus, the commutator is correctly written as the following form, which is similar to the K\"{a}ll\'{e}n-Lehmann representation:
\begin{align}
\lim_{ \epsilon \rightarrow 0} A_{\epsilon}(t_{12},\Omega_{12})
& \simeq -\left(d^{\Delta}\right)_{\infty} \frac{\ii}{2 } \frac{1}{\varepsilon^{2 (\Delta-d/2+1)}} 
\int_0^\infty d \mu 
\left( \mu^2 \right)^{\Delta-d/2}
\Delta(\mu; t,r),
\end{align}
where 
\begin{align}
\Delta(\mu; t,r)=\int_0^\infty d k
\sin\left( \sqrt{\mu^2+k^2} \, t \right)
\tilde{C}_{k/\varepsilon}(\cos(\varepsilon r))
\end{align}
The function $\Delta(\mu; t,r)$
is proportional to the commutator of the scalar field with a mass $\mu$
on Minkowski spacetime~$[\phi(x),\phi(y)]$ in which we identified $t=x_0-y_0$ and $r=\sqrt{(x_i-y_i)^2}$,
as shown in Section~\ref{sec:free-cft}.

Here, the integration over the weight, $\int_0^\infty d \mu 
\left( \mu^2 \right)^{\Delta-d/2}$, is divergent whereas it should be converged 
for
the K\"{a}ll\'{e}n-Lehmann representation of the commutator.
This is because the asymptotic fields (in the flat space case) can not 
be defined in the non-trivial CFT.
It also clearly be related to the fact that
the equal time commutator is not well-defined and 
we need the smearing of the local operators as we showed before.

Instead of the smearing, we can introduce the UV cut-off $\Lambda$ for the integral of $\mu$,
like $\int_0^\Lambda d \mu $, to make the integral converge.
However, this is not appropriate because this divergence does not due to the 
theory itself. The divergence appears because we consider the ``ill-defined'' operators,
i.e. the local operators. If we consider the well defined operators, which can be (spacetime)
smeared local operators, 
there are no divergence and there are no need of the cut-off to define the commutator, as we have seen.

\subsubsection{$\Delta=d/2$ case}

In general, $\left(d^{\Delta}\right)_{2n+l}^{l}$ is complicated and 
it is difficult to perform the summation over $n$ explicitly.
However, 
we can perform the summation when $\Delta=d/2$ where 
$\left(d^{\Delta}\right)_{2n+l}^{l}$ does not depend on $n$ and $l$: $\left(d^{\Delta=d/2}\right)_{2n+l}^{l} = \frac{2\pi^{d/2}}{\Gamma(d/2)}$.~\footnote{
If the CFT has the holographic dual, this $\Delta$ corresponds to the 
scalar in the bulk whose mass saturates the Breitenlohner-Freedman bound~\cite{Breitenlohner:1982jf}.}
For this case, the commutator is 
\begin{align}
A_{\epsilon}(t_{12},\Omega_{12})
& =
\frac{2\pi^{d/2}}{\Gamma(d/2)} \sum_{n,l=0}^{\infty}
e^{-(\Delta+2n+l)\epsilon}(-2\ii)\sin\left((\Delta+2n+l)t_{12}\right) \tilde{C}_{l}(\Omega_{12}) \\
& =
\frac{2\pi^{d/2}}{\Gamma(d/2)} \sum_{l=0}^{\infty}
\left(
\frac{ -e^{(\Delta+l)(-\epsilon+\ii t_{12})} }{1-e^{-2 \epsilon+2 \ii t_{12}} }
+\frac{ e^{(\Delta+l)(-\epsilon-\ii t_{12})} }{1-e^{-2 \epsilon-2 \ii t_{12}} }
\right) 
\tilde{C}_{l}(\Omega_{12}).
\end{align}
Thus, if $e^{2 \ii t_{12}} \neq 1$,
\begin{align}
\lim_{\epsilon \rightarrow 0}  A_{\epsilon}(t_{12},\Omega_{12})
& =
\frac{2\pi^{d/2}}{\Gamma(d/2)} \frac{1}{\ii \sin t_{12} }
\sum_{l=0}^{\infty}
\cos ( (d/2-1+l) t_{12} )
\tilde{C}_{l}(\Omega_{12}) \nonumber \\
& =
\frac{1}{2} \frac{1}{\sin t_{12} }
[\frac{d}{dt_{1}} \mathcal{O}_{\rm free}(t_{1},\Omega_{1}),  \mathcal{O}_{\rm free}(t_{2},\Omega_{2})]
,
\label{cd2}
\end{align}
where the expression is interpreted as a distribution in space
and $\mathcal{O}_{\rm free}(t,\Omega)$ is the free massless scalar considered in Section~\ref{sec:free-cft}.
Note that it becomes large when $t_{12}$ is small
because of the $\frac{1}{\sin t_{12} }$ factor.
This reflects the fact that the equal time commutator diverges.

Note also that this expression does not vanish at $t_{12}=0$
although we are considering the commutator of the same operators.
This is consistent because 
$\lim_{t_{12} \rightarrow 0} A_{\epsilon}(t_{12},\Omega_{12}) =0$
where $\epsilon$ was fixed at a finite value.

We find that 
the commutator of the scalar operators with $\Delta=d/2$  represented in the ``angular momentum'' eigen modes, 
$\mathcal{O}_{lm}(t) := \int d \Omega \, \mathcal{O}_L (t,\Omega) Y_{lm}(\Omega)$,
as
\begin{align}
\bra{0} [\mathcal{O}(t_1)_{lm}  ,\mathcal{O}(t_2)_{l' m'}  \ket{0}
& =
\frac{1}{\ii \sin t_{12} }
\cos ( (d/2-1+l) t_{12} )
\delta_{l,l'} \delta_{m.m'}
\label{cd21}
\end{align}
for $e^{2 \ii t_{12}} \neq 1$.

\section*{Acknowledgments}

L.N. would like to thank to Yukawa institute for Theoretical Physics at Kyoto University for hospitality. Discussions during the YITP atom-type visiting program were useful to proceed this work.
This work was supported by JSPS KAKENHI Grant Number 17K05414.

\vspace{1cm}

\noindent
{\bf Note added}: 

As this article was being completed, we received the preprint~\cite{deBoer}. In that paper, they 
discussed some general aspects of commutators of local operators in CFT from OPE.

\appendix
\section{Conventions and notations}\label{app:conventions}
\subsection{Spherical harmonics}\label{app:spherical}
In this article we use the convention for spherical harmonics as they satisfy
\begin{equation}
\int d\Omega
Y^{\text{(here)}}_{l,m}(\Omega)
Y^{\text{(here)}}_{l',m'}(\Omega)
=\delta_{l,l'}\delta_{m,m'}.
\end{equation}
On the other hand, another convention is used in~\cite{Terashima:2019wed}
where spherical harmonics satisfy
\begin{equation}
\frac{1}{\text{Area}(\mathbb{S}^{d-1})}
\int d\Omega
Y^{\text{(there)}}_{l,m}(\Omega)
Y^{\text{(there)}}_{l',m'}(\Omega)
=\delta_{l,l'}\delta_{m,m'} ,
\end{equation}
where 
\begin{equation}
\text{Area}(\mathbb{S}^{d-1})=\int d\Omega = \frac{2\pi^{d/2}}{\Gamma(d/2)} .
\end{equation}
which reduces $4\pi $ in $d = 3$.
%
The relation between them are given by
\begin{equation}\label{eq:spherical-harm-normalization}
Y_{l,m}^{(\text{here})}(\Omega)
=
\frac{1}{\sqrt{
\text{Area}(\mathbb{S}^{d-1})
}}\,
Y_{l,m}^{(\text{there})}(\Omega).
\end{equation}

%

\subsection{Gegenbauer polynomials and addition theorem}
\label{sec:gegenbauer}
In this paper, the Gegenbauer polynomials $C_{n}^{\alpha}$~\cite{Avery} is given by
\begin{align}
C_s^{\alpha} \left(\eta \right) 
=\sum_{p=0}^{[\frac12 s]} 
\frac{(-1)^p (2 \eta)^{s-2p} }{p! (s-2p)!}
{\Gamma(\alpha+s-p) \over \Gamma(\alpha)},
\end{align}
which reduces to the Legendre polynomial
for $\alpha=1/2$.
The Gegenbauer polynomials are normalized as 
\begin{equation}
\int
dx
(1-x^{2})^{\alpha-1/2}
\left[
C_{n}^{(\alpha)}
(x)
\right]^{2}
=
\frac{\pi 2^{1-2\alpha}\Gamma(n+2\alpha)}{n!(n+\alpha)\left[\Gamma(\alpha)\right]^{2}}.
\end{equation}

The spherical harmonics satisfy the following relation which is so-called the addition theorem
\begin{equation}
\sum_{m}
Y_{l,m}(\Omega_{1})
Y_{l,m}(\Omega_{2})
=
\frac{d+2 l-2  }{d-2} 
C^{(d/2-1)}_{l}(\Omega_{12}).
\end{equation}

\section{Approximation of $\left(d^{\Delta}\right)_{2n+l}^{l}$}
\label{sec:approx}
Stirling's formula is given by
\begin{equation}
\Gamma(z)\simeq
\sqrt{\frac{2\pi}{z}}
\left(
\frac{z}{e}
\right)^{z} ,\quad(\abs{\arg{z}}<\pi-\epsilon, \abs{z}\to\infty ) .
\end{equation}
Using this formlua and the following equation, 
\begin{equation}
\frac{(x+a)^{x+a}}{x^{x}}
\simeq
e^{a}x^{a} , \quad(x\to\infty) ,
\end{equation}
we have
\begin{align}
(x)_{a}&:=\frac{\Gamma(x+a)}{\Gamma(x)}
\\
&\simeq
\sqrt{\frac{x}{x+a}}
\frac{e^{x}}{e^{x+a}}
\frac{(x+a)^{x+a}}{x^{x}}
\\
&\simeq
e^{-a}
\frac{(x+a)^{x+a}}{x^{x}}
\\
&\simeq
e^{-a}e^{a}x^{a}=x^{a}, \quad(x\to\infty) .
\end{align}
Thus when $n,l$ are large we have
\begin{align}
\left(d^{\Delta}\right)_{2n+l}^{l}
&=
\frac{2\pi^{d/2}}{\Gamma(d/2)}
\frac{\Gamma(d/2)\Gamma(\Delta+n+l)\Gamma(\Delta+1-d/2+n)}{\Gamma(\Delta)\Gamma(\Delta+1-d/2)\Gamma(n+1)\Gamma(n+l+d/2)}
\\
&=
\frac{2\pi^{d/2}}{\Gamma(d/2)}
\frac{\Gamma(d/2)(n+1)_{\Delta-d/2}(\Delta+n+l)_{\Delta-d/2}}{\Gamma(\Delta)\Gamma(n+1+d/2)}
\\
&\simeq
\frac{2\pi^{d/2}}{\Gamma(d/2)}
\frac{\Gamma(d/2)}{\Gamma(\Delta)\Gamma(\Delta+1-d/2)}
\left((n+1)(n+l+d/2)\right)^{\Delta-d/2} 
\\
&\simeq
\frac{2\pi^{d/2}}{\Gamma(d/2)}
\frac{\Gamma(d/2)}{\Gamma(\Delta)\Gamma(\Delta+1-d/2)}
\left(n(n+l)\right)^{\Delta-d/2} .
\end{align}

\bibliographystyle{JHEP}

\end{document}